\documentstyle[12pt]{article}
\def\OldText#1\EndOldText{{\tiny {\bf OLD TEXT:} #1 {\bf END} }}

  \oddsidemargin=\evensidemargin
\let\a=\alpha \let\b=\beta  \let\d=\delta \let\e=\varepsilon
  \let\th=\theta  
    \let\p=\pi \let\r=\rho
\let\s=\sigma   \let\c=\chi
\let\ps=\psi 
\let\Ph=\phi     
   
\def\0{\over }    \def\1{\vec }   \def\2{{1\over2}} \def\3{{\ss}}
\def\4{{1\over4}} \def\5{\bar }   \def\6{\partial } \def\7#1{{#1}\llap{/}}
\def\8#1{{\textstyle{#1}}}        \def\9#1{{\bf {#1}}}
\def\_#1{$\underline{\hbox{#1}}$} \def\^#1{$\overline{\hbox{#1}}$}

\def\<{\langle } \def\>{\rangle }  
 
\def \({\left( } \def \){\right) }

 \let\eq=\equiv    \let\aus=\in
      \let\and=\wedge
     
\def\|#1{{}_{\bigg|_{#1}}}

\def\mao#1{\mathop{\rm #1}\nolimits}      \def\tr{\mao{tr}} 
  \def\mod{\mao{mod}}
\def\sign{\mao{sign}}

  \def\cg{{\cal G}} 
  \def\co{{\cal O}} \def\cp{{\cal P}}
   
\def\cR{{\cal R}}
\def\inbar{\vrule height1.5ex width.4pt depth0pt} %\font\ZZsf=cmss12
\def\ifundefined#1{\expandafter\ifx\csname#1\endcsname\relax}
\makeatletter \ifundefined{new@mathgroup} {} \else  \input{oldlfont.sty} \fi
\def\ZZ{\relax{\sf Z\kern-.4em \sf Z}}  \def\IR{\relax{\rm I\kern-.18em R}}
\def\IN{\relax{\rm I\kern-.18em N}} \def\IP{\relax{\rm I\kern-.18em P}}
\def\IQ{\relax\,\hbox{$\inbar\kern-.3em{\rm Q}$}}
\def\IC{\hbox{\,$\inbar\kern-.3em{\rm C}$}}
\def\citen#1{\if@filesw \immediate\write \@auxout {\string\citation{#1}}\fi%
\@tempcntb\m@ne \let\@h@ld\relax \def\@citea{}%
\@for \@citeb:=#1\do {\@ifundefined {b@\@citeb}%
    {\@h@ld\@citea\@tempcntb\m@ne{\bf ?}%
    \@warning {Citation `\@citeb ' on page \thepage \space undefined}}%
    {\@tempcnta\@tempcntb \advance\@tempcnta\@ne
    \setbox\z@\hbox\bgroup\ifcat0\csname b@\@citeb \endcsname \relax
       \egroup \@tempcntb\number\csname b@\@citeb \endcsname \relax
       \else \egroup \@tempcntb\m@ne \fi \ifnum\@tempcnta=\@tempcntb
       \ifx\@h@ld\relax \edef \@h@ld{\@citea\csname b@\@citeb\endcsname}%
       \else \edef\@h@ld{\hbox{--}\penalty\@highpenalty
	      \csname b@\@citeb\endcsname}\fi
    \else \@h@ld\@citea\csname b@\@citeb \endcsname \let\@h@ld\relax \fi}%
 \def\@citea{,\penalty\@highpenalty\hskip.13em plus.13em minus.13em}}\@h@ld}
\def\@citex[#1]#2{\@cite{\citen{#2}}{#1}}%
\def\@cite#1#2{\leavevmode\unskip
  \ifnum\lastpenalty=\z@\penalty\@highpenalty\fi% highpenalty before
  \ [{\multiply\@highpenalty 3 #1%              % triple-highpenalties within.
  \if@tempswa,\penalty\@highpenalty\ #2\fi}]}   % and before note.
\makeatother % END OF   "CITE.STY"   (\citen can be used for citation numbers)

\def\beq{\begin{equation}} \def\eeq{\end{equation}} \def\eql#1{\label{#1}\eeq}
\def\bea{\begin{eqnarray}} \def\eea{\end{eqnarray}} 
\def\fnote#1#2{\begingroup\def\thefootnote{#1}\footnote{#2}
	   \addtocounter{footnote}{-1}\endgroup}    \let\nn=\nonumber

\def\plb#1 #2 {Phys. Lett. {\bf B#1} #2 }
\def\phr#1 #2 {Phys. Rep. {\bf  #1} #2 } 
\def\npb#1 #2 {Nucl. Phys. {\bf B#1} #2 }
\def\aph#1 #2 {Ann. Phys. {\bf #1} #2 }  
\def\jmp#1 #2 {J. Math. Phys. {\bf #1} #2 }
\def\prd#1 #2 {Phys. Rev. {\bf D#1} #2 }
\def\prl#1 #2 {Phys. Rev. Lett. {\bf #1} #2 }
\def\rmp#1 #2 {Rev. Mod. Phys.  {\bf #1} #2 }
\def\zpc#1 #2 {Z. Phys. {\bf #1C} #2 }
\def\cmp#1 #2 {Comm. Math. Phys. {\bf #1} #2 }
\def\mpl#1 #2 {Mod. Phys. Lett. {\bf A#1} #2 }
\def\ijmp#1 #2 {Int. J. Mod. Phys. {\bf A#1} #2 }

\let\3=\ss \catcode`\"=\active \let"=\"

\let\msk=\medskip \def\ng{n_{27}} \def\na{n_{\overline{27}}} 

\begin{document}

\catcode`\@=11

\newif\if@fewtab\@fewtabtrue
%%%%%%%%%%%%%%%%%%%%%
%%%%% draftdate %%%%%
{\count255=\time\divide\count255 by 60
\xdef\hourmin{\number\count255}
\multiply\count255 by-60\advance\count255 by\time
\xdef\hourmin{\hourmin:\ifnum\count255<10 0\fi\the\count255}}
\def\ps@draft{\let\@mkboth\@gobbletwo
    \def\@oddhead{}
    \def\@oddfoot
%      {\hbox to 7 cm{$\scriptstyle Draft\ version:\ \draftdate$
       {\hbox to 7 cm{\tiny \versionno
       \hfil}\hskip -7cm\hfil\rm\thepage \hfil}
    \def\@evenhead{}\let\@evenfoot\@oddfoot}

\newcommand{\version}[1] {\typeout{}\typeout{#1}\typeout{}\vskip3mm
                   \centerline{{\tt DRAFT -- #1 -- \today}} \vskip3mm}

\def\draftdate{\number\month/\number\day/\number\year\ \ \ \hourmin }
\def\draft{\pagestyle{draft}\thispagestyle{draft}
\global\def\draftcontrol{1}}
\global\def\draftcontrol{0}
\catcode`\@=12
%%%%%%%%%%%%%%%%%%%%%%%%%%%%%%%%%%%%%%
% a few macros                      %%
%%%%%%%%%%%%%%%%%%%%%%%%%%%%%%%%%%%%%%

\def\bfe           {{\bf1}}
\def\bgg           {Bern\-stein\hy Gel\-fand\hy Gel\-fand}
\def\cdim          {conformal dimension}
\def\cft           {conformal field theory}
\def\cfts          {conformal field theories}
\def\complex       {{\dl C}}
\def\csa           {Cartan subalgebra}
\def\cwb           {Cartan\hy Weyl basis}
\def\cy            {Calabi\hy Yau}
\def\cym           {Calabi\hy Yau manifold}
\newcommand{\dfrac}[2]{{\displaystyle\frac{#1}{#2}}}
\long\def\del#1    \enddel{}
\long\def\drac#1#2{{\displaystyle\frac{#1}{#2}}}
\def\dyd           {Dynkin diagram}
\def\eE            {{\rm e}}
\def\emt           {energy-momentum tensor}
\def\epp          {extended Poin\-car\'e polynomial}
\newcommand{\erf}[1]{(\ref{#1})}
\def\furu          {fusion rule}
\def\hsc           {hermitian symmetric coset}
\newcommand{\hsp}[1] {\mbox{\hspace{#1 mm}}}
\def\hy            {$\mbox{-\hspace{-.66 mm}-}$}
\def\ii            {{\rm i}}
\newcommand{\internal}[1]{\hfill\break {\footnotesize #1} \hfill\break}
\def\kma           {Kac\hy Moo\-dy algebra}
\def\kpf           {Kac\hy Peterson formula}
\def\ks            {Ka\-za\-ma\hy Su\-zu\-ki}
\long\def\labl#1   {\label{#1}\ee\mbox{ }\\[-12 mm]\query{#1}\\[5 mm] }
\def\lg            {Landau\hy Ginz\-burg}
\def\lgo           {Landau\hy Ginz\-burg orbifold}
\def\lie           {Lie algebra}
\def\Lie           {Lie group}
\def\morE          {......... \fbox{more ?} ......... \query{\fbox{more ?}}}
\def\neW           {......... \fbox{new !} ......... \query{\fbox{new !}}}
\def\ngen          {\mbox{$N_{27}^{}$}}
\def\nagen         {\mbox{$N_{\overline{27}}$}}
\def\NN            {$N=2$}
\def\NS            {Neveu\hy Schwarz\ }
\def\one           {\mbox{\small $1\!\!$}1}
\def\onehalf       {\mbox{$\frac12$}}
\def\opa           {operator product algebra}
\def\opp          {ordinary Poin\-car\'e polynomial}
\def\pp           {Poin\-car\'e polynomial}
\def\ptx           {\mbox{${\cal P}(t,x)$}}
\long\def\query#1{\hskip 0pt{\vadjust{\everypar={}\small\vtop to 0pt{\hbox{}%
     \vskip -13pt\rlap{\hbox to 46.3pc{\hfil{\vtop{\hsize=8pc\tolerance=6000%
     \hfuzz=.5pc\rightskip=0pt plus 3em\noindent#1}}}}\vss}}}}%
\def\R             {Ramond }
\long\def\rank#1   {\mbox{rank}\,#1}
\def\reals         {\mbox{${\sf I}\!{\sf R}$}}
\def\rep           {representation}
\def\Rep           {Representation}
\def\rgs           {Ramond ground state}
\def\scft          {superconformal field theory}
\def\scfts         {superconformal field theories}
\def\sign          {\mbox{sign\,}}
\def\smat          {$S$-matrix}
      \newcommand{\sps}{\mbox{$J_s$}} \def\sp{s}
\def\suco          {superconformal}
      \newcommand{\sus}{\mbox{$J_v$}} \def\su{v}
\def\susy          {su\-per\-sym\-me\-try}
\def\twodim        {two-di\-men\-si\-o\-nal}
\def\ue            {\mbox{u(1)}}
\def\uE            {\mbox{u$_1$}}
\def\wrt           {with respect to\ }
\def\wzw           {WZW\ }
\def\WZW           {Wess\hy Zumino\hy Witten}
\def\wzwm          {WZW model}
\def\wzwt          {WZW theory}
\def\wzwts         {WZW theories}
\def\zet           {\ZZ}
\def\zetminuso     {\mbox{${\zet}_{\leq 0}$}}
\def\zetplus       {\mbox{${\zet}_{>0}$}}
\def\zetpluso      {\mbox{${\zet}_{\geq 0}$}}

%%%%%%%%%%%%%%%%%%%%%%%%%%%%%%%%%%%%%%
%%%%%%%%% end of macros      %%%%%%%%%
%%%%%%%%%%%%%%%%%%%%%%%%%%%%%%%%%%%%%%

%\draft
%\version\versionno

\def\nikhef{NIKHEF 95-006}
\def\tuw{TUW--95-03}
{\hfill\nikhef\vskip-9pt \hfill\tuw\vskip-9pt    \hfill hep-th/9503174}
%\vskip 15mm\centerline{\hss\Huge\bf Extending the Poincar\'e Polynomial\hss}
\vskip 15mm\centerline{\hss\Huge\bf On the extended Poincar\'e Polynomial\hss}
\begin{center} \vskip 8mm
       Maximilian KREUZER\fnote{*}
      {e-mail: kreuzer@tph16.tuwien.ac.at}
\vskip 3mm
       Institut f"ur Theoretische Physik, Technische Universit"at Wien,\\
       Wiedner Hauptstra\3e 8--10/136, A--1040 Vienna, AUSTRIA
%      CERN, Theory Division\\CH--1211 Geneva 23, SWITZERLAND
\vskip 6mm               and
\vskip 3mm
       Christoph SCHWEIGERT\fnote{\#}
      {e-mail: t10@nikhef.nl}
\vskip 3mm
       ~~ NIKHEF-H/FOM, ~~ P.O. Box 41882\\
       1009 DB Amsterdam,  THE NETHERLANDS

%\vfil
\vspace{15mm}                      {\bf ABSTRACT}                \end{center}

We show that the numbers of generations and anti-generations of a (2,2)
string compactification with diagonal internal theory can be expressed in
terms of certain specifications of the elliptic genus of the untwisted
internal theory which can be computed from the \pp. To establish this result
we show that there are no cancellations of positive and negative
contributions to the Euler characteristic within a fixed twisted sector.
For our considerations we recast the orbifolding procedure into an algebraic
language using simple currents.
Turning the argument around, this allows us to define the `\epp' $P(t,x)$,
which encodes information on the orbits of the spinor current under fusion,
for non-diagonal \NN\ \suco\ field theories.
As an application, we derive an explicit formula for $P(t,x)$ for general
Landau-Ginzburg orbifolds.

\thispagestyle{empty} %\vfil\noindent    \nikhef\\ \tuw\\ March 1995
\baselineskip=14pt \msk \newpage

\section{Introduction}

The most accessible and most important quantities characterizing a string
compactification arise in the description of the spectrum of massless
excitations.
A large part of this spectrum is topological: For a (2,2) vacuum the numbers
$\ng$ and $\na$ of generations and
anti-generations can be computed in terms of the charge degeneracies of the
Ramond ground states of the GSO-projected theory. In contrast, the numbers of
gauge singlets and extra gauge bosons require additional
information on the internal conformal field theory and may vary as we move in
moduli space. The charge degeneracies of \rgs s of an \NN\ theory are
conveniently encoded in their generating function, the \pp\
\beq
     P(t,\bar t)% = {\rm Tr}_{{\cal R}} t^{ J_0} \bar t^{ \bar J_0}
            = (t\bar t)^{c/6}{\rm Tr}_{{R_0}} t^{ J_0} \bar t^{ \bar J_0}\, ,
\eeq
where the trace extends over the set $R_0$ of Ramond ground states in the
theory. Note that, strictly speaking, $P$ is not a polynomial in $t$ and
$\bar t$,
since typically fractional powers occur in the normalization we have chosen.
In practice one frequently only knows the \pp\ for the theory
{\em before} GSO-projection; this polynomial is comparatively
easy to compute even for rather complicated $N=2$ theories like Kazama-Suzuki
models \cite{lvw,sche3,fuSc}. To compute $\ng$ and $\na$, it is necessary to
know the \pp\ of the GSO-projected theory as well.

In \cite{but} it was shown that the `Euler number'
$\c=2(\na-\ng)$ of a string vacuum that results from the GSO-projection of a
diagonal theory without additional twists can be obtained directly from the
\pp\ $P(t)=P(t,1)$ of the original internal \cft. To compute $\ng$ and $\na$
separately,
the \epp\ was introduced \cite{sche3} as an efficient bookkeeping device,
which encodes information about the intersection of the orbits of the spinor
current with the set of \rgs s.

It was observed \cite{sche3,fuSc} that in all known cases the \epp s of two
\suco\ field theories  are identical whenever the \pp s are the same. This
suggests that the \epp\ may be fixed by the \pp\ alone, and that this holds
independently of the underlying $N=2$ theory.
In particular, the numbers of generations and anti-generations might be
computable directly from the \pp. In the present paper we show that for
diagonal theories this
is indeed the case. For a fixed twist of a diagonal theory,
all Ramond ground states with the same right-moving $U(1)$ charge contribute
with the same sign to the Euler number. As a consequence, we can generalize
the Buturovi\'c formula and explicitly calculate $\ng$ and $\na$ separately.
This, in turn, tightly constrains the \epp.

In Section 2 we recall some basic facts about the implementation of the
generalized GSO-projection using simple current modular invariants \cite{sy}.
This construction is closely related to Gepner's `shift vector'
method \cite{gep} and to Vafa's orbifolding procedure \cite{v}. In a brief
digression we discuss the modding of `geometrical' and `quantum' symmetries
from an algebraic point of view. The definition of the \epp\ and its
use for the calculation of the non-singlet spectrum is a simple application
of these ideas.

In Section 3 we analyse the index used by Buturovi\'c for the calculation
of the Euler number.
We show that the contributions coming from a fixed twisted sector never
cancel. This result is used to derive formulae for $\ng$ and $\na$.

The translation of the definition of the \epp\ into the orbifold language
allows to generalize it to arbitrary $N=2$ \suco\ field theories. It is,
therefore, straightforward to derive an explicit formula for the \epp\ for
orbifolds of \lg\ models; this is done in Section 4. It is crucial that this
works for \cfts\ with arbitrary central charge $c$. This has
important practical consequences, since the \epp\ encodes all information
that is relevant for the topological part of the spectrum of any vacuum that
is built from a tensor product. Hence, if we reproduce the \epp\ of, say,
a Kazama--Suzuki model with some \lgo, then we know that all possible
spectra of tensor products containing this model must be identical to the
spectrum of the tensor product that is obtained by replacing the
Kazama-Suzuki model in this tensor product by the \lgo. As an example we
consider the relation of an infinite series of non-Hermitian coset models
to \lgo s.

\section{Simple currents and $\bf N=2$ \suco\ theories}

To obtain a consistent string theory from an \NN\ \suco\ theory, it is
necessary to implement a number of projections, which can be described by
means of integer
spin simple currents. We will therefore first recall some basic facts about
simple currents, and describe the ones that are present in any \NN\ theory and
that constitute the `generic center'. For more details we refer the reader to
the review \cite{sc90r}.

A simple current $J$ is a primary field whose fusion product with any
primary field $\Ph$ contains a single primary field $J\Ph$  with multiplicity
one \cite{sc90r}. The set of primary fields therefore decomposes into orbits
with respect to the fusion product $(J)^n\Ph$ whose maximal length $N$ is
called
the order of the simple current; the set of all simple currents of a rational
\cft\ forms a finite abelian group,  the {\em center}. For any primary field
$\Phi_i$ we define the monodromy charge by $Q_J(\Ph_i)\eq h_i+h_J-h_{J i}$,
where $h_i$ is the conformal weight of the primary field $\Ph_i$ and $h_J$ is
the conformal weight of the simple current $J$. In operator products this
charge is conserved modulo \zet.  Thus the phase $\exp(2\p\ii Q_J)$ is
conserved under the operator product, which implies that
the center acts as a discrete symmetry group on the \cft. (Of course, not all
symmetries have to be generated by simple currents and there are symmetries
of a \cft\ which are not described by its center.)
The monodromy charges and conformal weights modulo \zet\
of the simple currents can be parametrized
by a matrix $R$ as
\beq
     R_{ij}= r_{ij}/N_i\eq Q_i(J_j) =  Q_j(J_i), ~~~~~
     h_{[\a]}\eq \2\sum_i r_{ii}\a^i-\2\sum_{ij}\a^iR_{ij}\a^j~~\mod~1,
\eeq
where $[\a]=\prod J_i^{\a_i}$, $N_i$ is the order of the current $J_i$,
and the diagonal elements $R_{ii}$ are defined modulo 2.

\subsection{Simple current modular invariants}

Simple currents can be used to construct modular invariants, which are
closely
related to orbifolding with respect to the corresponding discrete symmetries.
To construct a modular invariant we need to choose a subgroup of the center
for which all diagonal entries in the corresponding monodromy matrix $r$ are
even, so that the
spin multiplied by the order is an integer (this condition is non-trivial
only for simple currents with even order; it is related to the level matching
condition in
the orbifold framework). From the definition of the monodromy charge it
follows that
\beq
     h([\a] Ph)\eq h(\Ph)+h([\a])-\a^iQ_i(\Ph)  \mod \zet  \, ,
\eql h
while the fact that the monodromy charge is conserved modulo integers implies
that
\beq
     Q_i([\a] Ph)\eq Q_i(\Ph)+R_{ij}\a^j  \mod \zet\ \, .
\eql Q
Furthermore, the $S$ matrix elements for fields that are on the
same orbits are related by phases,
\beq
     S_{[\a]a,[\b]b}=S_{a,b}e^{2\p \ii(\a^kQ_k(b)+\b^kQ_k(b)+\a^k R_{kl}\a^l)}
\eeq
(the indices $k$ and $l$ are to be summed over the chosen subgroup
of the center). It can now be checked that the matrix
\beq
     M_{\Ph,[\a]\Ph}={\rm Mult}(\Ph)\prod_i\d_\zet\(Q_i(\Ph)+ X_{ij}\a^j\)
\eql X
commutes with the generators $S$ and $T$ of modular transformations, if $X$ is
properly quantized and $X+X^T\eq R$ modulo \zet;
$\d_\zet(r)$ is 1 if $r\in\zet$
and 0 otherwise. ${\rm Mult}(\Ph)$ is the multiplicity of the primary field
$\Ph$, i.e.\ the ratio of the size of the subgroup of the center that defines
the modular invariant over the size of the orbit containing $\Ph$.
It can be shown, using certain regularity assumptions, that (\ref X) is the
most general simple current modular invariant, i.e.\ a modular invariant that
only relates primary
fields on the same orbits of the center \cite{sci,gs}.  Note that the freedom
in the choice
of the anti-symmetric part of $X$ corresponds to the freedom in the choice of
phases of the projections in the twisted sectors of orbifolds, which are
called discrete torsions.\footnote{
	Because of the coincidence of the resulting string vacua, simple
	current modular invariants and orbifolds with discrete torsion
	were first conjectured to be equivalent in ref. \cite{fiqs}.}
The vector $\a$ of exponents can be interpreted as
a twist, i.e. it tells us which sector of the orbifold a certain non-diagonal
field comes from (this interpretation is consistent with the twist selection
rules). It is interesting to remark that there are several infinite series of
simple current modular invariants known \cite{fusS} which are not of the form
\erf X, and therefore do not possess a description in terms of
orbifolds with discrete torsion;  however, these modular invariants do not
give rise to consistent \cfts.

\subsection{Simple currents of $N=2$ \suco\ theories}

Returning to \suco\ field theory,
the existence of a simple current with conformal weight $3/2$ and order 2, the
supercurrent $J_\su$, follows already from $N=1$ \susy; its
monodromy charge is $0$ for NS states and $1/2$ in the Ramond sector.
   Note that we are using the word supercurrent to denote a primary field
   $J_\su$. However, it can happen that one can construct more than one
   supercharge out of descendants of this primary field.\footnote{
   This is the case
   e.g.\ in \NN\ coset models \cite{kasu} where both supercharges belong
   to the same primary field which has a representative with trivial quantum
   numbers except for the ${\rm so(}d)$ part, where it is equal to the vector
   $v$.}
For $N=2$ \suco\ models we have at least one more simple current, namely
the \rgs\ $J_\sp$ of highest $U(1)$ charge, which in our normalization is
$c/6$. This simple current, to which we will refer as the
{\em spinor current}, implements the spectral flow: In terms of the
bosonized $U(1)$ current of the \NN\ algebra $J(z)=\sqrt{c\03}\6X(z)$ it is
given by
$J_\sp = \exp{\ii\sqrt{3\0c}X}$. It follows from the operator product of the
free boson $X$ that the monodromy charge $Q_\sp$ of the spinor current
is related to the $U(1)$ charge $Q$ by $Q_\sp\equiv-Q/2$ modulo 1.
Thus, if $1/M$ is the charge quantum in the NS sector and $\hat c=c/3=k/M$,
then the order of the spinor current $J_{\sp}$ is $2M$ if $k$ is even and
$4M$ if $k$ is odd, because the charges in the Ramond sector are shifted by
$c/6$. If $k$ is odd, then $J_{\su}=(J_{\sp})^{2M}$, so that the order of the
`generic' center, which is present in any rational \NN\ theory,
is $4M$ in both cases.
$J_{\sp}$ is a Ramond ground state, hence $h_{\sp}=c/24$ and
$Q_{\sp}(J_{\sp})=-c/12$.
Putting the pieces together we find for the matrix of monodromies
\beq
     R_{\su,\su}=0, ~~~~~ R_{\su,\sp}=1/2, ~~~~~
     R_{\sp,\sp}=n-c/12~~~\hbox{with }n=
	\cases{k/2&~$k$ even\msk\cr~1&~$k$ odd.}
\eeq
In the case of \nn\ minimal models $k$ is the level and $M=k+2$.

A consistent heterotic (or type II) vacuum is now obtained by the
following procedure. Working in the bosonic framework (i.e. after the
bosonic string map) we have to tensor the internal $c=9$ $N=2$ \suco\ theories
with a $D_5$ Kac--Moody algebra at level 1
(we omit a factor of $E_8$ at level 1,
which is irrelevant for our considerations). For having a well-defined
supersymmetry generator
it is essential that we project out all mixed states, which do not have all
factors in the NS sector or all in the Ramond sector, from the tensor product.
Furthermore, the spectrum of the string should be space-time supersymmetric.

Both requirements can be implemented with simple currents:
Note that the total spinor current $J_s^{tot}:=s\otimes J_s$, i.e.\
the product of the spinor of $D_5$ with the spinor current $J_s$ of the
internal $c=9$ theory, and the product $J_v^{tot}=v\otimes J_v$ of the
vector $v$ of $D_5$ with the supercurrent $J_v$
both have integral spin since $h(J_s)=c/24=3/8$ and
$h(s)=5/8$. Moreover, all monodromies vanish. Hence, if we
choose no torsion between $J_s^{tot}$ and $J_v^{tot}$, i.e. $X=0$ in
eq.~(\ref X), then all fields with non-integral monodromy charges or,
equivalently,
U(1) charges that are not even\footnote{
     	We have to project on {\em even} charges since
	we work after having applied the bosonic string map.}
are projected out and both  simple currents extend the chiral algebra of the
theory. This is exactly what we want to achieve, since $J_s^{tot}$ can be
combined with right moving bosons to yield the gravitino vertex; on the gauge
side of the heterotic string it extends the gauge group from $D_5$ to $E_6$.
Primary fields with mixed boundary conditions have
half-integral monodromy charge with respect to $J_v^{tot}$; if the internal
theory is itself a tensor product of $N=2$ theories, then the group generated
by all bilinears in the vector currents has to extend the chiral algebra
in order to align Ramond states and NS states (see \cite{sche3,schW2}).

\subsection{The extended \pp}

We are now in a position to define the \epp.
Since the non-singlet $E_6$ representations all come from
Ramond ground states of the GSO-projected theory, we encode, as for the
ordinary \pp, only information on these states and their charges.
Both the alignment of boundary conditions and the generalized GSO-projection
correspond to integral spin
simple
currents and we may, in a first step, disregard
the projections corresponding to the
product of $\d$ functions in the expression (\ref X) and consider the
`unprojected orbifolds'. Eventually, to obtain the projected orbifold, we
just have to omit the
contributions with non-integral monodromy charges. The aim of the
\epp\ is to encode all information about an \NN\ \suco\ theory which is
necessary
to compute the massless spectrum of any tensor product with $c=9$ that
contains
this model as one factor. To this end we also need to encode information on
the twists: We first define the `full extended' \pp{}
\beq 								\label{fepp}
     \cp(t,\5t,x,\s)=\sum_{l\ge0}~\sum_{k=0}^1x^l\s^kP_{l,k}(t,\5t),
\eeq
where $P_{l,k}(t,\5t)$ is the \pp\ of the unprojected sector twisted by
$J_s^{2l}J_v^k$. Hence, $P_{l,k}$ is obtained by looking for all pairs
$(\a,\a')$ of Ramond ground states with $\a'=J_s^{2l}J_v^k\a$; the charges of
$\a$ and $\a'$ are encoded in the exponents of $t$ and $\5t$, respectively.

The information on the location of \rgs s on the simple current orbits of
$J_s$ and $J_v$ is important if
we consider tensor products of $N=2$ factor theories.
For simplicity we restrict ourselves to the case that the tensor product
contains only two factors; what we are really interested in in this situation
is the modular invariant obtained for the total spinor
current $J_s^{(1)}J_s^{(2)}$ after the alignment of R and NS sectors. For the
tensor product we thus obtain the `full extended' \pp{} by the prescription
\beq
     \cp(t,\5t,x,\s)=\sum_{l\ge0} x^l \(
	\sum_{k=0}^1P^{(1)}_{l,k}(t,\5t)P^{(2)}_{l,k}(t,\5t)+
	\s~\sum_{k=0}^1P^{(1)}_{l,k}(t,\5t)P^{(2)}_{l,1-k}(t,\5t)\).
\eeq
Hence, (\ref{fepp}) indeed encodes all the information from a factor theory
that enters the computation of the generation numbers in arbitrary tensor
products. In fact, this information is still redundant: Consider a pair of
R ground states $(\a,\a')$ whose contribution to $P_{l,k}$ is
$t^{Q(\alpha)+{c\06}}\,\5t^{Q(\alpha')+{c\06}}$. Then eq.\ (\ref Q) implies
that
\beq
     Q(\a')\equiv-2Q_s(\a')\equiv Q(\a)-2\(2lR_{ss}+kR_{sv}\)
	\equiv Q(\a)+l\8{c\03}-k ~~~ \mod ~ 2.
\eeq
Hence the exponent
\beq
     k\eq Q(\a)+l{c\03}-Q(\a') ~~~\mod ~ 2\label{sign}
\eeq
of $\s$ is fixed in terms of the other exponents.
So we can set $\s$ to $-1$;
the negative sign is a convenient choice because a twist
by an odd number of supercurrents $J_v$ implies a negative contribution to
the index.

In the original definition of the \epp\ \cite{sche3} Schellekens, in addition,
put
$\5t=1$. If we are only interested in applications to heterotic (2,2) string
vacua built from diagonal theories, this is still a sufficient amount of
information for the following reason. We can turn the above argument
around and conclude that for given exponents of $t$, $x$ and $\s$ the charge
$Q(\a')$ of $\a'$ is known modulo 2. For a symmetric (2,2) vacuum this is
all we need to know. In \cite{sche3} it was implicitly assumed that, for a
twist with a given power of the spinor current, all contributions with a
fixed charge $Q(\alpha)$ contribute
with the same sign to the index. In the next section we will show that this
is  the case for all diagonal \NN\ \suco\ field theories. Therefore, for this
application, the \epp
\beq
     P(t,x):=\cp(t,1,x,-1)
\eeq
indeed encodes sufficient information for the calculation of the massless
non-singlet modes for arbitrary tensor products (without additional twists).

It should be clear how the simple current construction of string vacua
is related to the orbi\-fold technique of ref. \cite v.
In that paper the theory is modded by the symmetry
$j=\exp (2\p \ii J_0)$, where $J_0$ is the zero mode of the left-moving U(1)
charge, in order to project to integral charges. The arguments
concerning modular invariance are `modulo GSO-projection', which means up to
half-integral contributions to conformal weights in the Neveu-Schwarz sector
and without specification
of the action of the symmetry in the Ramond sector, i.e. up to a possible
twist by the supercurrent. Our discussion shows that, in the specific
situation which the \epp\ was invented for, the \pp\ $P(t,\5t)$ for the
projected total internal \cft\ and $P(t,x)$ encode equivalent information.
In a more general situation, $P(t,\5t,x)=\cp(t,\5t,x,-1)$ would be more
appropriate, since the complete information about left- and right-charges
allows to treat non-diagonal theories, whereas the information on the twist,
encoded by the exponent of $x$, allows to compute the relevant information
individually for each factor in a tensor product.

\subsection{Orbifolds and chiral algebras}

At this place we want to make a few general remarks on orbifolds and their
description in an algebraic framework. In the cases of orbifolds we considered
so far,  the chiral algebra was always {\em extended} by some integer spin
simple currents (in case of discrete torsion $X\neq X^T$ the left and right
extension can be different \cite{sci}). There is, however, in the algebraic
approach (compare e.g.\ \cite{dvvv}) a different use of the word `orbifold'
which
denotes the case when one {\em restricts} the chiral algebra to some
subalgebra of
the original chiral algebra.
This subalgebra has to contain the Virasoro algebra of the original theory;
hence the conformal anomaly has the same value $c$ in both \cfts.
This is the case e.g.\ for
the $\zet_2$ orbifold of a free boson $X$ compactified on a circle: while the
original algebra contains all polynomials in $\partial X$, the chiral algebra
of the orbifold only contains even polynomials in $\partial X$, which are
invariant under $\partial X \mapsto -\partial X$.

In the case of a general orbifold, however, the situation is much more
involved: It may be that the new chiral algebra ${\cal A}'$ neither
contains the original chiral algebra ${\cal A}$ nor is itself contained
in ${\cal A}$; however, in any case the intersection ${\cal A}'\cap{\cal A}$
is non-empty and contains the Virasoro algebra of ${\cal A}$. It may even be
that there is no change in the chiral algebra at all and that the invariant
is a pure
automorphism invariant. This can be observed rather explicitly with the simple
current modular invariants if the simple current has non-integer spin. An
example are the D-type modular invariants of su(2) at level $k = 2 \bmod 4$,
which correspond to the automorphism of the fusion rule which maps the
primary field
with Dynkin label $l$ onto itself if $l$ is even and to $k-l$ if $l$ is odd.

Note that any abelian orbifold has a symmetry group that is isomorphic to
the twists defining the orbifold. It was observed for simple current
invariants \cite{sc90r}
that, in case of a cyclic center, the square of a modular invariant gives
back the diagonal one if the modular invariant does
not extend the algebra. In the orbifold language this corresponds to the
modding by the `quantum symmetry' (i.e. the symmetry implied by the twist
selection rule), which also gives back the original theory. Since the latter
does not refer to the chiral algebra at all it is natural to ask if we can
return to the original theory by such a modding also if the chiral algebra
of a simple current modular invariant is extended. In that case we must not
consider the maximal chiral algebra, since then different twists contribute
to the same primary field and we cannot see the twist selection rule (this
implies that, in such a situation, we must work with operator products
instead of fusion rules, which are only well-defined with respect to the
maximal chiral algebra). Here we are indeed in the situation discussed in
\cite{dvvv}:   The smaller algebra typically has more irreducible
representations than the original one, and these must provide the twisted
fields. By modding combinations of `classical' and `quantum' symmetries in an
orbifold it is clear that we can have a mixed situation, where the chiral
algebra is restricted and then re-extended by some fields from the twisted
sectors.\footnote{
     Technically, the quantum symmetries can be implemented by introducing
     twists that act trivially on the original CFT, except for certain
     discrete torsions with the twists defining the orbifold \cite{odt}.}
It would be interesting to extend this picture of going back and forth
between the original CFT and the orbifold to non-abelian twists.

\section{The massless spectrum}

\subsection{A non-cancellation theorem}

We are now going to prove that for a fixed twist with respect to the spinor
current $J_s$ two \rgs s with the same charge can contribute to the index
only with the same sign, in other words, that there are no cancellations
within one fixed twisted sector. This result will allow us in the next
subsection to compute $\na$ and $\ng$ separately.

To prove this statement we fix a twist $J_s^{2l}$ and assume that there are
two \rgs s $|\a\>$ and $|\b\>$
with the same $U(1)$ charge $Q$ which
both contribute to the \epp\ in that sector, but with different signs. We
will show that this leads to a contradiction.

After possibly interchanging the role of $|\a\>$ and $|\b\>$, we may assume
that $|\a'\>=(J_vJ_s^{2l})_0|\a\>$ and $|\b'\>=(J_s^{2l})_0|\b\>$ both are
non-vanishing. The index 0 indicates the zero modes of the operators, and
hence
$|\a'\>$ as well as  $|\b'\>$ are \rgs s as well.
Since the superpartner of a \rgs\ is not a \rgs, but rather has conformal
weight $h > \frac c{24}$ we conclude that $(J_s^{2l})_0|\a\>=0$ has to
vanish.

The main tool to derive a contradiction is an operator whose zero mode
connects $|\a\>$ and $|\b\>$.
It can be constructed as follows: consider the primary
fields $\co_\a$ and $\co_\b$ that generate $|\a\>$ and $|\b\>$ from the
vacuum. In any \NN\ \suco\ field theory  we can split off a U(1) factor which
is generated by the U(1) current of the \NN\ algebra and write the theory as
a tensor product of the U(1) theory and some theory which only contains
uncharged operators, with a non-product modular invariant. Therefore
we can write $\co_\a=\eE^{\ii Q\sqrt{3\0c}X}\hat\co_\a$ and
$\co_\b=\eE^{\ii Q\sqrt{3\0c}X}\hat\co_\b$ in terms of the bosonized $U(1)$
current and neutral operators $\hat\co$. These operators can a priori contain
a polynomial in the derivative of $X$ which is uncharged as well. However, the
derivative of $X$ is proportional to the U(1) current
in the \NN\ algebra, which is an element of the chiral algebra. Hence the
presence of such operators in $\hat\co$ would contradict the requirement that
both $|\a\>$ and $|\b\>$ are primary.

If we now use spectral flow to relate
$\co_\b$ to a chiral operator $\co_c$ and the conjugate operator
$\co_{\a^c}$ to an anti-chiral operator $\co_a$, we observe that
the zero mode of $\co_a$ sends $|\a\>$ into the R ground state
$|J_s^{-1}\>$ with lowest $U(1)$ charge and that the zero mode
of $\co_c$ sends that state into $|\b\>$. Hence $\b\>=\ps_0|\a\>$ with
$\ps_0=(\co_c)_0(\co_a)_0$. Since $\ps_0$ does not change the $X$ dependent
part of $\co_\a$ it has to be uncharged and can therefore contain at most
a polynomial in $\partial X$  and its derivatives,
which acts on $|\a\>$ as a c-number $y$. Since
the only property of $\ps_0$ we are interested in is that it maps $|\a\>$ on
$\b\>$ we can replace the polynomial in $\partial X$ by $y$ and assume
that $\ps_0$ is chosen to be completely
independent of $X$ and its derivatives, and thus commutes with $J_s^{2l}$
which is entirely build out of exponentials of $X$.

Putting the pieces together we find $|\b'\>=(J_s^{2l})_0\ps_0|\a\>
=\ps_0(J_s^{2l})_0|\a\>=0$,
which is a contradiction to our original assumption that $\a$ and $\b$
both contribute to the index with the same $J_s^{2l}$ twist but with different
signs.

\subsection{Computation of the massless spectrum}

In \cite{but} is was shown how to compute the index of the GSO-projected
theory directly from the Poincar\'e polynomial of the untwisted theory. To
this end, the following quantities have been introduced:
\beq
     \cp_{r,s} :=\tr_{R_0^{(s)}}\eE^{2\p \ii r (J_0-{c\06})}e^{\ii\p(Q_+-Q_-)}
            =\tr_{R^{(s)}}e^{\ii\p  (r c/3 +Q_+-Q_-)}\eE^{2\p \ii r J_0}
             q^{L_0-c/24}\5q^{\5L_0-c/24}.
%   \tr_{R_0^{(s)}}(-1)^{rc/3}e^{2\p ir J_0}e^{i\p  (Q_+-Q_-)}
%            =\tr_{R^{(s)}}\eE^{i\p  (r c/3 +Q_+-Q_-)}\eE^{2\p ir J_0},
\label{butin} \eeq
Here $\tr_{R_0^{(s)}}$ denotes the trace over all \rgs s in the $s$-th twisted
sector and $\tr_{R^{(s)}}$ the trace over the whole $s$-th twisted Ramond
sector.
$Q_\pm$ denotes the left and the right moving U(1) charge, respectively.
The projection that is implemented by inserting the sum over $r$ of
$(\exp(2\p \ii (J_0-c/6)))^r$ along the second cohomology cycle of the torus
takes into account the shift of the charges by $c/6$ under spectral flow to
the Ramond sector.

The numbers $\cp_{r,s}$ are index-like quantities and close relatives
of the elliptic genus \cite{scwa4,witt17}. To make the relation precise let us
define for any twisted sector $s$  the following trace:
\beq
Z_s (q,r;\bar q, \bar r) := \tr_{R^{(s)}} (-1)^F q^{L_0-c/24} \bar
q^{\5L_0-c/24} \eE^{2\pi \ii (r J_0 - \bar r \bar J_0)} \label{trace} \eeq
Our notation slightly differs from
the usual definition by the factors of $2\pi$ and a relative minus sign for
$J$ and
$\bar J$. The elliptic genus can be obtained from this expression by setting
$\bar r =0$, while $r$ can take any value. Using index arguments one finds
that this function does not depend on $\bar q$ any more and one ends up with
a character valued index which is a signed sum over the characters of \rgs s.

There is however, another set of index-like quantities, which can be obtained
by restricting
the values of $r$ and $\bar r$ in a different manner: Both are required to
be integers.
The usual index argument shows that for integral values of $r$ and $\bar r$
the contributions
of two superpartners to $Z_s$ cancel and that therefore $Z_s$ does not depend
on $q$ and $\bar q$ any more.

The numbers $\cp_{r,s}$ can be obtained from \erf{trace} by setting
$\bar r$ to zero,
\beq
     \cp_{r,s} = Z_s(r,0) \eE^{\pi \ii rc/3}  \,  ;
\eeq
as we have set $\bar r =0$ this is equal to the elliptic genus, which at
values of $r$ we have chosen does not depend on $q$ either and therefore
simply is a number. Since these quantities are indices they are modular
invariant and
we can express those belonging to the twisted sectors using only expressions
in the
untwisted theory \cite{but}: $\cp_{r,s}=\cp_{dr+bs,cr+as}=\cp_{r\cap s,0}$,
where $r\cap s$
denotes the greatest common divisor of $r$ and $s$.

In \cite{but} it was argued that the contribution $\c_s$ of the $s$-th
twisted sector to the Euler number is
\beq
     \c_s={1\0M}\sum_{r=0}^{M-1}\cp_{r\cap s}, ~~~~ \mbox{where}\,
     \cp_r:=\cp_{r,0}=P(t=\eE^{2\p \ii r},\5t=1). \label{seceul}
\eeq
Any of these numbers must be an integer; we will use this later to derive
restrictions on the possible form of the \epp. The Euler number itself is
given by
\beq        \c = \sum_{s=0}^{M-1} \c_s   .  \eeq
Since we know that within a fixed twisted sector the contributions to $\c_s$
do not cancel, we can  also derive an expression for the total number
$\Sigma=|\cR|$ of \rgs s in the GSO-projected theory in terms of the
\pp\ of the untwisted theory:
\beq
     \Sigma=\tr_{R_0}\bfe =\sum_s \left| \c_s\right| =
     \sum_s\left|{1\0M}\sum_r\cp_{r,s}\right| \, .
\eeq

If we assume that space-time supersymmetry is not extended,
we have the two relations
\beq \c = 2(\ng-\na) \quad\mbox{and}\qquad \Sigma= 4+2(\ng+\na) \, \eeq
which determine both $\ng$ and $\na$ separately. This assumption is no real
obstacle in practice since space-time supersymmetry can only be extended if
the Euler characteristic $\chi$ vanishes. More precisely, the \pp\ has to
factorize into a $c=3$ and $c=6$ part, and the inverse charge quantum $M$
must be the (co-prime) product of the inverse charge quanta of the factors.

\subsection{Constraints on the indices $\cp_r$}

The indices $\cp_r$ are highly constrained by consistency
requirements.  First of all,  they have all to be integers, since
$\cp_s = \cp_{0,s}$ counts the number of \rgs s in the $s$-th twisted
sector. In addition, the twist is $\bmod M$ and hence $\cp_{r,s}$ is periodic
$\bmod M$ in the second label. Hence we have
\beq \cp_r = \cp_{r,0} = \cp_{r,M} = \cp_{r\cap M}\, ,  \eeq
and $\cp_r$ can  depend only on $r\cap M$. As a consequence, also the
contribution of the $s$-th twisted sector to the Euler number only depends on
$s\cap M$.

In addition, there are divisibility constraints coming from formula
\erf{seceul}
for the Euler number in the twisted sectors. If, e.g., $p$ is a
prime divisor of $M$, then $\c_p = \frac1p(\cp_p +(p-1)\cp_1)$ must be an
integer
and hence $\cp_p-\cp_1$ must be a multiple of $p$.

We can also restrict the charges that occur in the $s$-th twisted sector:
To begin with, assume that $s$ and $M$ are co-prime; then we have also that
$(r\cap s )\cap M=1$ for all $r$. We can use \erf{seceul} to calculate
\beq \c_s = \frac1M \sum_{r=0}^{M-1} \cp_{r\cap s} = \cp_1=\cp_s=\cp_{0,s}.
\label{calcu}\eeq
Note that the sum over $r$ implements on the states in the Ramond sector a
projection on charges for which $Q+\frac c6 \in\zet$. Since there are no
cancellations in $\c_s$, the calculation shows
that this projection keeps all states in the $s$-th twisted sector
and that therefore the charges of the Ramond ground states in this twisted
sector obey $Q+\frac c6 \in\zet$.
It is straightforward to generalize this constraint to the case when
$s$ and $M$ are not co-prime, $s\cap M = l$. A calculation analogous to
\erf{calcu} shows that in this case
\beq
     \frac lM \sum_{r=0}^{M/l-1} \cp_{lr,s} = \cp_{0,s},
\eeq
which shows that for the Ramond ground states in
these sectors the charges obey $Q+\frac c6 \in \frac 1l \zet$.

So far we have assumed that $\hat c:=c/3$
is an integer. This condition is not really a restriction, because we can
always fulfill it for a tensor product with a suitable
number of minimal models:
For $N=2$ theories $k$ has to be even if $M$ is even \cite{sum}. Hence
we get an integral total $\hat c$ if we tensor with $n$ minimal models at
levels $k_i=M-2$, where $n=k/2$ if $k$ is even and $n=(k+M)/2$ if $k$ is odd.
The \epp\ for minimal models
\beq
     P(x,t^M)^{({\rm mm})}=\sum_{s=1}^{M-1}t^{s-1}{1-(-)^sx^{s}\01-(-)^Mx^{M}}
\eql{eppmm}
was derived in \cite{sche3}. Note that there is a single contribution
to the $s^{th}$ twisted sector which has charge $(s-1)/M$. Thus, for
a central charge $c/3=k/M\not\in\ZZ$, the charges of Ramond ground states
in the unprojected twisted sectors fulfill
\beq
     Q+{c\06}\in -n{s-1\0M}+{\ZZ\0s\cap M}, ~~~~~
     n\equiv k(M+1)/2~\mod~M.
\eeq
The \pp\ of a minimal model evaluated at $e^{2\p\ii\,r/M}$ is
$-e^{-2\p\ii\,r/M}$. Therefore $\cp_r'=e^{-2\p\ii\,nr/M}\cp_r$ must be
an integer, which, up to a sign, counts the numbers of Ramond ground states
in the respective sector.

For the problem of constructing $P(t,x)$ from $P(t)$ this means that, due
to the non-cancellation theorem, we have restricted the possible \epp s for
a given \pp\ to a
finite set which is rather small.
All remaining freedom is in the integral
part of the charges (or up to multiples of $1/l$ if $l=s\cap M>1$).
In many cases the remaining freedom can be fixed by various consistency
requirements. To illustrate this, we consider the $E_7$ invariant\footnote{
	This is not a diagonal conformal field theory, but all contributions
	to the \pp\ are diagonal. Our proof of non-cancellation
	can be extended to this situation.}
of the \NN\ minimal model at level 16, which has the \pp\
\beq
     P(t^9)=1+t^2+t^3+t^4+t^5+t^6+t^8.
\eeq
We find $\cp'_1=1$ and $\cp'_3=-2$. Since the central charge $c$ is smaller
than 3, all charges are smaller than 1 and the \epp\ must be of the form
\beq
     P(x,t^9)=P(t^9)+x+x^2t^5-x^3(at+bt^4+ct^7)+x^4t^6+x^{5}t^2
             -x^{6}(ct+bt^4+at^7)+x^{7}t^3+x^{8}t^8 \, .
\eql{E7}
$\cp'_3=-2$ implies that $a+b+c=2$; note that charge conjugation relates
the sectors with $s=3$ and $s=6$.
The projection to `integral charges' (after tensoring 4
minimal models
at level 7)
acts non-trivially
only in the sectors 3 and 6, where it keeps $(\cp'_1+\cp'_1+\cp'_3)/3=0$
states. This implies $a=0$, since $a$ is the coefficient
of the only term which contributes to states with an integer charge in
the tensor product.
With a simple
observation we can also fix the coefficients $b$ and $c$: If we tensor our
model with itself
and in addition with 8 minimal models at level 7,
then not only $(at)^2$, but also $2(bt^4)(ct^7)$ has the
correct charge to survive the projection in the $3^{rd}$ sector.
This time the projection keeps
$((\cp'_1)^2+(\cp'_1)^2+(\cp'_3)^2)/3=2$ states, which implies that $2bc=2$.
Since $b+c=2$ we conclude $b=c=1$.

If $c$ is larger than 3 then we are (almost) always left with some ambiguity,
because all information we obtained so far is insensitive to the integral
part of all $U(1)$ charges. In that case we can, however, use the
sum rule for $U(1)$ charges that was derived in \cite{sum} to
constrain the integral parts of the charges (by applying this sum rule to
tensor products we can derive further constraints, which was sufficient
to fix all remaining freedom in a number of cases
that we considered with $c>3$). In \cite{sche3,fuSc} consistency constraints
for tensor products were also used to fix some ambiguity in the \epp\ that
arose from field identification fixed point problems in Kazama--Suzuki
coset models. Note, however, that
the only information on the specific \cft\ we used was the \opp.

\section{Extended \pp s for \lgo s}

In this section we  discuss how to compute the \epp\ in the orbifold
framework and, as an application, derive an explicit formulae for the \epp s
of \lg\ models and
their orbifolds. Recall that the operator $j=\exp(2\p \ii J_0)$
generates the symmetry group that leaves all states with integral
charges invariant and that $j^M$ is the identity in the NS sector.
We denote by $P_l(t,\5t)$ the \pp\ in the unprojected
sector twisted by $j^l$ (up to superpartners, as discussed in section 2.3).
In general $P_l(t,\5t)$ will be
asymmetric in $t$ and $\5t$, even if the untwisted theory was diagonal.
Up to a sign, the $l$-th twisted sector contributes to the \epp\ with
$ x^{l} P_l(t,1)$.
If the left and right charges $Q_\pm$ of the states in the twisted sectors
are known at least modulo two, we can use Equation \erf{sign} to compute
this sign as $(-1)^s$, with $ s=Q_+-Q_-+r\,\8{c/3}$.
This expression for $s$ must be an integer even before the projection to
invariant states.  Hence we obtain the formula
\beq
     P(x,t)=\sum_{r\ge0}(\eE^{\ii\p c/3}x)^{r}
            P_r(\eE^{\ii\p}t,\eE^{-\ii\p})
\eeq
for the \epp.

\subsection{Untwisted \lg\ models}

\def\ttb{(t\bar t)}
For \lg\ models, we are now in position to compute the \epp:
The charges of the twisted `vacua' in the Ramond sector are known
\cite{v,iv,mmlg}
to be
\beq
     Q_\pm|j^l\>_R=\Big(\sum_{lq_i\in\ZZ}(q_i-\2)
        \pm\sum_{lq_i\not\in\ZZ}(\th_i^{(l)}-\2)\Big)|j^l\>_R
\eeq
where $q_i=n_i/M$ is the $U(1)$ charge of the $i^{th}$ field and
$\th_i^{(l)}=lq_i-[lq_i]$ with $[x]$ denoting the greatest integer
smaller
than $x$. Thus the \pp\ in the $l^{th}$ sector is {\em before} projection:
\beq
     P_l(t,\bar t)=\prod_{lq_i\not\in\ZZ}t^{\th_i^{(l)}-q_i}
        \bar t^{1-\th_i^{(l)}-q_i}  %\ttb^{\2-q_i}(t/\bar t)^{\th_i^{(l)}-\2}
        \prod_{lq_i\in\ZZ}{1-\ttb^{1-q_i}\01-\ttb^{q_i}},
\eeq
the last factor comes from the chiral fields that are invariant under the
twist and therefore contribute to Ramond ground states in the twisted sectors.
Since $c/3=k/M=\sum (1-2q_i)$ we obtain for the \epp
\beq
      P(x,t)={1\01-(-)^kx^{M} }%^{-1}
     \sum_{r=0}^{M-1}(-)^{s_r}x^{r}  \prod_{rq_i\not\in\ZZ}t^{\th_i^{(r)}-q_i}
       \prod_{rq_i\in\ZZ}{t^{1-q_i}-1\0t^{q_i}-1},
\label{30}\eeq
where $s_r\eq Q_+-Q_-+r c/3\eq r N- N_{tw}(r)$,
$N$ is the total number of fields
and $N_{tw}(r)$ is the number of fields that are not invariant under $j^r$.

Applying this formula to the \NN\ minimal models with the diagonal modular
invariant we recover the expression
\erf{eppmm}; for the $E_7$ invariant we find agreement with \erf{E7}.
We also checked formula \erf{30} for a large subclass of Grassmannian coset
models, the $CP_n$ models, for which a \lg\ description is known \cite{lvw}.
These coset theories correspond to Hermitian symmetric spaces of the form
\beq
     A(n,1,k)={{\rm su}(n+1)_k\0{\rm su}(n)\oplus{\rm u}(1)}
\eeq
with central charge $ c={3nk\0k+n+1}$.
For our purpose the relevant data are the $U(1)$ charges
$q_i={i\0m+n+1}$ of the chiral superfields with $1\le i\le m$.
Note that the \nn\ minimal model at level $k$ can be described as a $CP_1$
model at level $k$.

\subsection{\lgo s}

For LG orbifolds we can proceed analogously and use the more general formula
%\ni$P(t,\5t)=\Tr_{(c,c)}t^{dJ_0}\5t^{d\5J_0}
%            =(t\5t)^{dc\06}\Tr_Rt^{dJ_0}\5t^{d\5J_0}$,
\beq
     Q_\pm|h\>_R=\Bigg(\pm\sum_{\th_i^h\not\in\ZZ}(\th_i^h-\2)+
     \sum_{\th_i^h\in\ZZ}(q_i-\2)\Bigg)|h\>_R
\eeq
for the charges of the vacua in the sectors twisted by group elements $h$.
Here $h$ acts on the fields $X_i$ like
$hX_i=\exp(2\p \ii\th_i^h)X_i$ with phases $0\le\th_i^h<1$.
In order to compute the \epp\ for a LG orbifold with twist group $\cg$ we have
to twist by all products of group elements $h\aus \cg$ with powers of $j$.
The projection, however, keeps all states that are invariant under the
elements of the
centralizer of $h$ in $\cg$, regardless of their charge (note that $j$
commutes with all linear symmetries of the LG potential, so the centralizer
is always independent of the $j$ twist).
The action of a group element $g$ that commutes with $h$ on the twisted
ground states in the Ramond sector can be shown to be
\beq
     g|h\>_R= (-)^{K_g(1+K_h)}\e(g,h)(\det g_{|_h})|h\>_R,
\eql{ghR}
where the phases $\e(g,h)$ fulfill the usual constraints on discrete torsions
\cite{iv}. For the concept of the \epp\ to make
sense we have to make sure that the supercurrent survives the
projection, i.e. all twists in $\cg$ must satisfy
$
     \deg g=(-1)^{K_g}
$
(see \cite{iv} for details). This fixes the group actions in the $R$ sector,
whose signs are parametrized by $(-1)^{K_g}$, and restricts the determinants
of allowed twists to real values.

Using the above formula for the charges of the twisted vacua $|h\>_R$
and the fact that the chiral excitations are described by
an effective LG theory consisting of the untwisted fields we find
\beq
     P_h(t,\bar t)=\prod_{\th_i^h>0}t^{\th_i^{h}-q_i}
        \bar t^{1-\th_i^{h}-q_i}  %\ttb^{\2-q_i}(t/\bar t)^{\th_i^{(l)}-\2}
        \prod_{\th_i^h=0}{1-\ttb^{1-q_i}\01-\ttb^{q_i}}
\eeq
for the unprojected contribution of the $h$ twisted sector to the \pp.
In order to obtain the \epp\ we have to sum over all twists
$h=\hat h j^r\in\cg\times\ZZ_M$;
the Ramond ground states in such a sector
contribute to the coefficient of $x^r$. Then we need to project
to states that are invariant under the group elements of $\cg$.

For abelian groups this projection can be implemented directly in the above
expression for $P_h$ in a convenient and efficient way:
Note that the denominators $1/(1-\ttb^{q_i})$ describe
the charge degeneracies of the free polynomial algebra and that the factors
$1-\ttb^{1-q_i}$ subtract the contributions from the ideals that are
generated by the gradients of the potential. It is essential that these
gradients are independent, which guarantees the correct counting and hence
that the complete expression is a polynomial.
Therefore we can implement the group transformation in a diagonal basis
$gX_i=\r_iX_i$ by replacing the denominators by $1/(1-\r_i\ttb^{q_i})$ and
the factors in the numerator by $1-\r_i^{-1}\ttb^{1-q_i}$. Additional phases
come from the transformation properties \erf{ghR} of the twisted vacua.
Eventually we can implement the projection by summing over $g\in\cg$.
It is, however, rather unpleasant to have polynomials with non-integral
coefficient in the denominator. This can be avoided if we write %replace
$1/(1-\r_i\ttb^{q_i})$ as %by
$\(\sum_{n=0}^{|\cg|-1}(\r_i\ttb^{q_i})^n\)/(1-\ttb^{|\cg|q_i})$, where we
used that $\r_i^{|\cg|}=1$.
(Instead of summing over the group it is  more efficient to work with formal
variables  describing the group action and to keep only
those terms in the numerator that are invariant under all generators of
the twist group $\cg$).

As a simple example we consider the $\ZZ_{2M}$ orbifold of the tensor product
of two minimal models at levels $k_i=2M-2$. In order to have a real
determinant,
thus keeping spectral flow and supersymmetry, the two factors should transform
with opposite phases. It is easy to see that this orbifold has the $U(1)$
charges quantized in units of $1/M$. Straightforward evaluation of the above
formulas yields the \epp\\
\vbox{
\bea
     &&\!\!\!\!\!\!P(x,t^M)={1\01-x^M}\Bigg({1-t^{2M-1}\01-t}+(2M-1)t^{M-1}\\
        &&\hspace{109pt}\nonumber
      +\sum_{r=1}^{M-1}x^r\((2r-1)t^{r-1}+(2M-2r-1)t^{M+r-1}\)\Bigg).
\eea}
The choice of this example is motivated by the observation that, after
tensoring
with a third minimal model at level $M-2$, the \lgo\
\beq \underbrace{X^{2M} +Y^{2M} }{/\ZZ_{2M}} + Z^{M} \eeq
reproduces the \pp\ of the non-hermitian symmetric coset model
\beq \frac{ (C_2)_{2M-3} \oplus so(6)_1}{(C_1)_{2M-2}\oplus{\rm u}(1)_{4M}} \,
{}.
\eeq
In fact, we find that the orbifold and the coset model also have the
same \epp\ and hence yield the same spectra in tensor products. Nonetheless,
the two theories cannot be isomorphic, because e.g.\ the numbers of simple
currents are different: The coset has $16M$ simple currents
while a single minimal model at level $k$ already has $4(k+2)$ simple
currents.

\section{Conclusions}

In this letter we have presented several new results concerning the
computation of the topological part of the massless spectrum of a string
compactification. For diagonal models we have proven a non-cancellation
theorem which allows us to compute the number of generations and
anti-generations
using the \pp\ only.

Furthermore, we have generalized the definition of the \epp\ for non-diagonal
theories and derived explicit formulae for it in the case of \lg\ models and
their orbifolds. Since the \epp\ is independent of the moduli, two
theories with the
same \epp\ need not be isomorphic. But at least we know that all the spectra
of all tensor products which contain them as factors are the same
(provided that no additional symmetries, which exist for only
one of the two models with coinciding \pp s, are modded).

If the left and right charges of all Ramond ground states in the twisted
sectors are known, as is the case for \lgo s, we can extend the definition
of the \pp\ to include also that information:
\beq
     P(x,t,\5t)=\sum_{r\ge0} (\eE^{\ii\p c/3}x)^{r}
P_r(\eE^{\ii\p}t,\eE^{-\ii\p}\5t) \, , \eeq
where the information of the sign now becomes redundant. This could be
useful for more general string vacua, as well as for other reasons, e.g.\
the comparison of  \NN\ theories that are given in a different formulation.

\vbox{\noindent {\bf Acknowledgements:}

We would like to thank J.\ Fuchs and A.N.\ Schellekens for helpful
discussions. One of us (C.S.) would like to thank the Institut f\"ur
Theoretische Physik der Technischen Universit"at Wien, where this work
started, and both of us thank the theory division at CERN, where this work
was finished, for
hospitality. Partial support by the {\it "Osterreichische Nationalbank}
under grant No. 5026 is gratefully acknowledged.
}

\end{document}